\documentclass[prl,twocolumn,showpacs]{revtex4}%
\usepackage{graphicx}
\usepackage{amsmath}
\usepackage{amsfonts}
\usepackage{amssymb}%
\begin{document}
\title{Nonlinear Dynamics and Chaos in Two Coupled Nanomechanical Resonators}
\author{R. B. Karabalin}
\affiliation{Department of Physics and Kavli Nanoscience Institute, California Institute of
Technology, Pasadena CA 91125}
\author{M. C. Cross}
\affiliation{Department of Physics and Kavli Nanoscience Institute, California Institute of
Technology, Pasadena CA 91125}
\author{M. L. Roukes}
\affiliation{Department of Physics and Kavli Nanoscience Institute, California Institute of
Technology, Pasadena CA 91125}
\date{\today}

\begin{abstract}
Two elastically coupled nanomechanical resonators driven independently near
their resonance frequencies show intricate nonlinear dynamics. The dynamics
provide a scheme for realizing a nanomechanical system with tunable frequency
and nonlinear properties. For large vibration amplitudes the system develops
spontaneous oscillations of amplitude modulation that also show period
doubling transitions and chaos. The complex nonlinear dynamics are
quantitatively predicted by a simple theoretical model.

\end{abstract}

\maketitle

Resonant nanoelectromechanical systems (NEMS) \cite{PhWorld} are
attracting interest in a broad variety of research areas and for many
possible applications due to their remarkable combination of
properties: small mass, high operating frequency, large quality factor,
and easily accessible nonlinearity \cite{LCreview}. Further development of NEMS
applications such as superior mass \cite{Phil}, force
\cite{RugarSpin} and charge \cite{ClelandCov} sensors, or reaching
the quantum limit of detection in mechanical systems \cite{Matt},
requires addressing several important challenges. For example, the
nonlinearity of the devices has to be either minimized or utilized
for improving performance \cite{DynamicRange}. In addition, the
large-scale integration of nanodevices demands a detailed
understanding of the behavior of coupled devices in NEMS arrays \cite{BuksRoukes,Lifshitz,Parametric,Eyal}.

In this paper we demonstrate complex nonlinear behavior of a pair of coupled
nanomechanical devices, and show that this can be quantitatively understood
from the basic physics of the devices. We show that the linear and weakly
nonlinear response of one oscillation can be modified by driving the second
oscillation, and, for some ranges of parameters of the devices, that the
linear response range of the first oscillation can be significantly extended.
When both oscillations are driven into their strongly nonlinear range more
complicated frequency-sweep response curves are found, corresponding to the
well known bistability of driven anharmonic \textquotedblleft
Duffing\textquotedblright\ resonators, but now with switching between a
variety of different stable states of the coupled pair. Spontaneous amplitude
modulation oscillations may develop, with frequencies characteristic of the
dissipation rates rather than of the intrinsic frequencies or their sums and
differences. These amplitude modulations show period doubling bifurcations and
chaos. The complex dynamics are reproduced quantitatively by a simple
theoretical model, giving us confidence that the nonlinear behavior of coupled
nanomechanical devices can be understood and controlled.

We study a system of two strongly coupled nonlinear nanoelectromechanical
resonators using a structure of doubly-clamped beams with a shared mechanical
ledge shown in Fig.~1a. The devices consist of a stack of three layers of
gallium arsenide (GaAs): a 100nm highly n-doped layer, a 50nm insulating
layer, and another 50nm layer that is highly p-doped. The piezoelectric
property of GaAs results in a highly efficient integrated actuation mechanism
described in \cite{D-NEMS}. A preliminary 120nm deep etch step is done to
isolate the actuation electrodes of the two beams, so that the two beams can
be addressed separately while retaining strong elastic coupling. Optical
interferometry is used for the motion transduction \cite{Carr}. The laser beam
is adjusted so that both beams are in the illuminated spot.

We model the behavior of the two strongly interacting nonlinear resonators by
a system of coupled equations of motion for the beam displacements
$x_{1},x_{2}$ in their fundamental modes
\begin{subequations}
\label{Eq_EOM}%
\begin{align}
\ddot{x}_{1}+\gamma_{1}\dot{x}_{1}+\omega_{1}^{2}x_{1}+\alpha_{1}x_{1}%
^{3}+D(x_{1}-x_{2})  &  =g_{D1}(t),\label{Eq_EOM_1}\\
\ddot{x}_{2}+\gamma_{2}\dot{x}_{2}+\omega_{2}^{2}x_{2}+\alpha_{2}x_{2}%
^{3}+D(x_{2}-x_{1})  &  =g_{D2}(t). \label{Eq_EOM_2}%
\end{align}
As well as the usual terms describing the resonant frequencies,
damping, and Duffing nonlinearity, we include a linear coupling term
in the displacements of strength $D$. The terms on the right hand
side are the external drives applied to the two beams, which are
controlled independently. The linear terms in the
equations, ignoring for now the drive and dissipation, give two modes
with frequencies $\omega_{I}$ and $\omega_{II}$ and corresponding
eigenvectors $\mathbf{e}_{I},\mathbf{e}_{II}$ \cite{Goldstein}. The
frequency difference of the modes results from both the intrinsic
frequency difference $\omega _{1}-\omega_{2}$ and the coupling $D$.

\begin{figure}
[tbh]
\begin{center}
\includegraphics[
height=1.8489in,
width=2.4093in
]%
{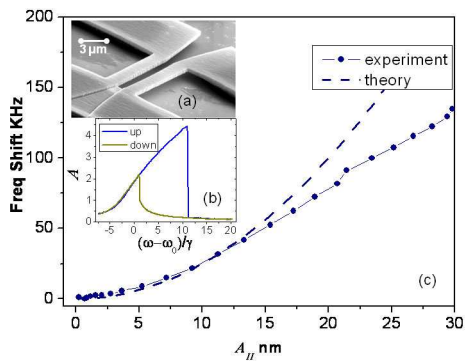}%
\caption{(color online) (a) SEM image of the system (beam dimensions: 6$\mu$m
x 500nm x 200nm). (b) Up (dark, blue) and down (light, yellow) frequency
sweeps of the amplitude $|A|$ of the response for a single drive of strength
4.3 times critical as a function of the frequency relative to the linear
resonance frequency $\omega_{0}$ and scaled by the width $\gamma$. The
amplitude is plotted in units of the maximum amplitude at the critical drive
strength. (c) Frequency shift of the weakly driven first mode as a function of
the displacement amplitude $|A_{II}|$ of the strongly driven second mode:
points and solid line -- experiment; dashed line -- small amplitude theory.}%
\label{Fig_1}%
\end{center}
\end{figure}
To investigate the behavior of the system, the two beams are
connected to two different sources with independent frequencies and
amplitudes. The monitored output variables are the amplitudes and
phases of a linear combination of the mechanical displacements of the
two beams at or near the two drive frequencies \footnote{The
combination measured depends on the geometry of the optical spot
relative to the beams, which we determine from the linear experiments
using a variety of drive combinations.}. By applying various
combinations of small amplitude signals to the two beams at
frequencies near the mode resonances, the linear coupling parameters
can be determined. For the particular device
shown in Fig.~1a the mode frequencies are determined to be $\omega_{I}%
/2\pi=16.79$ MHz and $\omega_{I}/2\pi=17.25$ MHz and the eigenvectors
$\mathbf{e}_{I}=(0.854,0.521)$ and $\mathbf{e}_{II}=(-0.521,0.854)$. The
frequencies of the individual resonators determined from inverting the mode
equations are $\omega_{1}/2\pi=16.71$ MHz and $\omega_{2}/2\pi=16.92$ MHz. The
frequency separation of 200 kHz is consistent with the fabrication tolerance.
The coupling strength $\sqrt{D}/2\pi=2.63$ MHz is consistent with the strength
of the elastic coupling found by finite element simulations of similar devices
\cite{RassulThesis}. Transforming the measured width of the modes back to the
original equations determines the values of $\gamma_{1},\gamma_{2}$. The
nonlinear parameters $\alpha_{1},\alpha_{2}$ are deduced from the expression
for the geometric nonlinearity \cite{DynamicRange} using the beam thickness
known from the fabrication and lengths calculated from the beam frequencies
$\omega_{1},\omega_{2}$ and the material constants \footnote{The structure of
the dynamical system depends on the ratio $\alpha_{I}/\alpha_{II}$ which is
close to unity for our nearly identical beams. The absolute magnitude of
$\alpha$ (approximately 0.0072 (MHz/nm)$^{2}$)
calibrates the amplitude of the response in nm.}.

The response of the system driven near resonance and for small dissipation and
driving can be calculated from (\ref{Eq_EOM}) using the standard methods of
secular perturbation theory \cite{LCreview}. This approach has previously used
for the case of parametrically driven nanomechanical devices
\cite{Lifshitz,Parametric,Eyal}, where hysteretic switches between different stable states
in frequency sweeps were also predicted. Briefly, we introduce the slowly
varying complex mode amplitudes $A_{I},A_{II}$ and forces $F_{I},F_{II}$ using
$x_{[I]}=\operatorname{Re}(A_{[I]}e^{i\omega_{\lbrack I]}t})$ and
$g_{D[I]}=\operatorname{Re}(F_{[I]}(t)e^{i\omega_{\lbrack I]}t})$ (where $[I]$
stands for either $I$ or $II$), substitute into the equations of motion, and
retain only the near resonant terms. This reduces the equations of motion to%
\end{subequations}
\begin{equation}
2i\omega_{I}\dot{A}_{I}+i\omega_{I}\gamma_{I}A_{I}+\alpha_{I}|A_{I}|^{2}%
A_{I}+\beta_{I}|A_{II}|^{2}A_{I}=F_{I}(t), \label{Eq_Amplitudes}%
\end{equation}
and a corresponding equation for $A_{II}$. The mode nonlinearity parameters
$\alpha_{\lbrack I]}$ and $\beta_{\lbrack I]}$ are calculated from $\alpha
_{1},\alpha_{2}$ and the eigenvectors $\mathbf{e}_{I},\mathbf{e}_{II}$ so that
all the parameters in (\ref{Eq_Amplitudes}) are known from linear measurements
and the beam geometry and material constants.

We first look at the case where each mode responds at the drive frequency
$\omega_{D[I]}$ which is set near the resonant frequency $\omega_{\lbrack I]}$
so that $A_{[I]}\propto e^{i(\omega_{D[I]}-\omega_{\lbrack I]})t}$, with%
\begin{equation}
|A_{I}|^{2}=\frac{|F_{I}|^{2}}{\left[  2\omega_{I}(\omega_{DI}-\omega
_{I})-\alpha_{I}|A_{I}|^{2}-\beta_{I}|A_{II}|^{2}\right]  ^{2}+\omega_{I}%
^{2}\gamma_{I}^{2}}, \label{Eq_Solutions}%
\end{equation}
etc. For a single drive (e.g.\ $A_{II}=F_{II}=0$), so that the cross-mode
nonlinear coupling proportional to $\beta_{{[I]}}$ is not involved, this
expression reproduces the regular Duffing response curve \cite{Nayfeh}.
Prominent features are the shift of the frequency of the maximum response to
larger values for positive $\alpha$ (nonlinear spring stiffening), and
bistability and hysteresis that develop above a critical drive strength. An
experimental example of upward and downward frequency sweeps for a drive
strength 4.3 times critical is shown in Fig.~1b.

\begin{figure}
[tbh]
\begin{center}
\includegraphics[
height=1.379in,
width=3.3723in
]%
{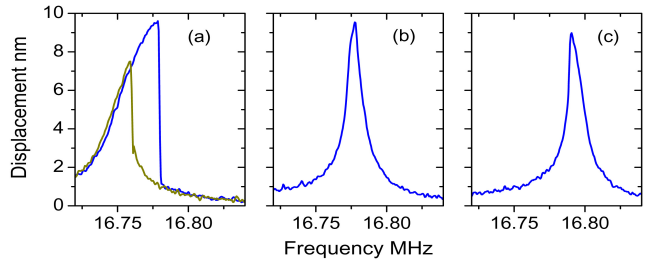}%
\caption{(color online) Frequency sweeps of the first mode response for
increasing amplitudes of the second mode: the dark (blue) lines are for upward
sweeps; the light (yellow) line in (a) is a downward sweep, showing the
hysteresis and amplitude jumps in this case. There are no hysteretic jumps for
(b) and (c), and so the downward sweeps are not shown. }%
\label{Fig_2}%
\end{center}
\end{figure}
We first
look at the linear and weakly nonlinear response of the first mode when the
second mode is driven into its strongly nonlinear regime. Retaining the
leading order effect of the first mode on the intensity of the second mode we
can write the numerator in (\ref{Eq_Solutions}) as%
\begin{equation}
\left[  2\omega_{I}(\omega_{DI}-\omega_{I})-\beta_{I}|A_{II}^{(0)}|^{2}%
-\bar{\alpha}_{I}|A_{I}|^{2}\right]  ^{2}+\omega_{I}^{2}\gamma_{I}%
^{2},\label{Eq_AI}%
\end{equation}
where $A_{II}^{(0)}$ is the solution for the second mode in the absence of
mode I given by the equation for $A_{II}$ corresponding to (\ref{Eq_Solutions}%
) assuming zero $A_{I}$. The new parameter $\bar{\alpha}_{I}$ is an effective
Duffing nonlinearity coefficient and is given by
$\bar{\alpha}_{I}=\alpha_{I}-2\beta_{I}\beta_{II}\omega_{II}\,\partial
|A_{II}^{(0)}|^{2}/\partial\omega_{DII}$.
Equation (\ref{Eq_AI}) predicts two important
effects: the frequency tuning of the first mode proportional to the
square of the amplitude of the second mode (upwards for positive
$\beta_{I}$); and the change in the effective nonlinear coefficient
for the motion of the first mode depending on the excitation strength
of the second mode through the last term in $\bar{\alpha}_{I}$.

To test the frequency tuning we excite the second mode at a drive level
approximately 4.3 times the critical value so that the spectral response is
the strongly nonlinear Duffing curve (see Fig.~1b). As the actuation frequency
of the second mode is steadily increased in small steps its vibration
amplitude rises over a wide frequency range until it drops to the lower
amplitude state beyond the maximum. The evolution of the spectral response of
the first mode is monitored at a driving level approximately four times lower
than the critical value for this mode using a network analyzer. The dependence
of the first mode frequency shift on the vibration amplitude of the second
mode is shown in Fig.~1c. The experimental results for frequency tuning
closely follow the predicted parabolic dependence $\beta_{I}\left\vert
A_{II}\right\vert ^{2}$ for amplitudes up to about 20nm. Monitoring the
frequency shift of one mode proportional to the square of the displacement of
a second mode has been proposed for quantum nondemolition measurements in
nanoelectromechanical systems \cite{Braginsky, PhysTodayQM}.

If the actuation level of the first mode is increased above the onset of
nonlinearity, while the second mode drive is kept at a much higher level, then
the effective nonlinear coefficient $\bar{\alpha}_{I}$ is decreased.
The decrease is largest when the second mode is driven on the portion of the
Duffing response curve where the intensity is increasing linearly with the
drive frequency $\left\vert A_{II}\right\vert ^{2}\simeq2\omega_{II}%
(\omega_{DII}-\omega_{{II}})/\alpha_{II}$. This gives the minimum effective
nonlinear coefficient $\bar{\alpha}_{I,\text{min}}=\alpha_{I}\left(
1-\frac{\beta_{I}\beta_{II}}{\alpha_{I}\alpha_{II}}\right)  $. This result
indicates that if the coupling is strong enough, so that $\beta_{I}\beta
_{II}>\alpha_{I}\alpha_{II}$, the minimum value of the effective nonlinear
coefficient is negative. In this case the resonance curve tilts to the left,
as opposed to the usual case for a doubly clamped beam where the peak leans to
the right. It also means that the nonlinear coefficient vanishes for some
drive strength and frequency of the second mode.

\begin{figure}
[tbh]
\begin{center}
\includegraphics[
height=2.3661in,
width=3.3723in
]%
{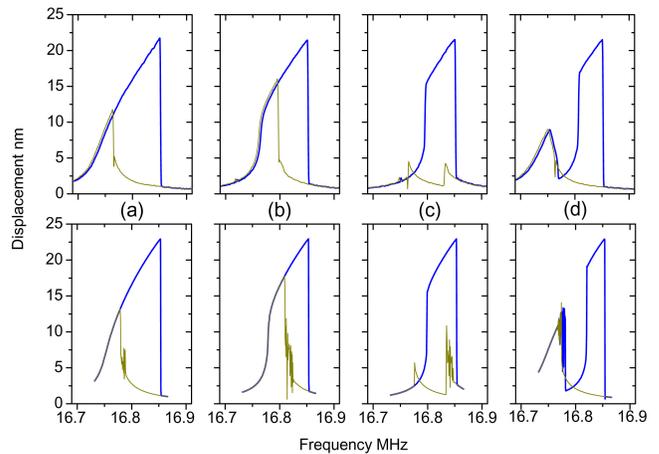}%
\caption{(color online) First mode frequency response as in Fig.~2 for
increasing values of the second mode drive frequency but for stronger driving
of the first mode (both modes are driven at about four times the critical
strength). The top plots are experimental measurements while the bottom ones
are theoretical simulations. The second mode response was monitored, but is
not shown.}%
\label{Fig_3}%
\end{center}
\end{figure}
Some experimental results for the first mode driven at twice the
critical strength illustrating this effect are shown in Fig.~2. The
plots show the shape of the resonance peak for three values of the
drive frequency of the second mode. In panel (a), the amplitude
$A_{II}$ of the second mode is low and the first mode spectral
response has the regular nonlinear Duffing shape leaning to the
right. For larger $A_{II}$ as in (b), the first mode resonance peak
shape assumes a form close to a Lorentzian with little nonlinearity
apparent. For even larger $A_{II}$ as in (c) the sign of the
effective Duffing coefficient becomes slightly negative, causing the
spectral response peak to lean to the left. The quenching of the
nonlinearity in (b) could be used to enhance the limited dynamic
range \cite{DynamicRange} of nanomechanical devices.

If the drive level of the first mode is increased further
a variety of new effects can be observed. Under these
conditions the dynamics of the system is not fully explained by the steady
state solutions as in (\ref{Eq_Solutions}), and so we solve for the expected
behavior by numerically integrating the time-dependent coupled equations
(\ref{Eq_Amplitudes}). Now the behavior is more complex, as the response of
the first mode becomes large enough to cause transitions in the second mode
response through the nonlinear coupling. As a result the spectral response
curves acquire peculiar nontrivial shapes, as shown in Fig.~3.
Numerical simulations of the equations give good predictions for the complex phenomena.

An obvious difference between the experimental and theoretical plots
in Fig.~3 is the noisy regions on the theoretical curves near the up
and down transitions. This difference actually results from the
different ways the plots are generated in theory and experiment, and
a more careful investigations shows consistent and interesting
dynamics in both experiment and theory: for the
drive frequencies near the transition points the fast (about 17MHz)
oscillating response becomes amplitude modulated with a frequency of
about 10 to 20 kHz \footnote{The theoretical points, which are the
end values of a long numerical simulation, depend on the phase of the
amplitude modulation at the end of each run, whereas this modulation
leads to a drop in the experimentally measured amplitude since
RF power is transferred outside the measurement bandwidth.}.
The theoretical equations (\ref{Eq_Amplitudes})
show that the frequency for the amplitude modulation is determined by
the line width, which in our experimental setup is about 8 kHz, and
is not related to sum and difference
frequencies of the two modes.%

\begin{figure}
[tbh]
\begin{center}
\includegraphics[
height=2.694in,
width=3.3723in
]%
{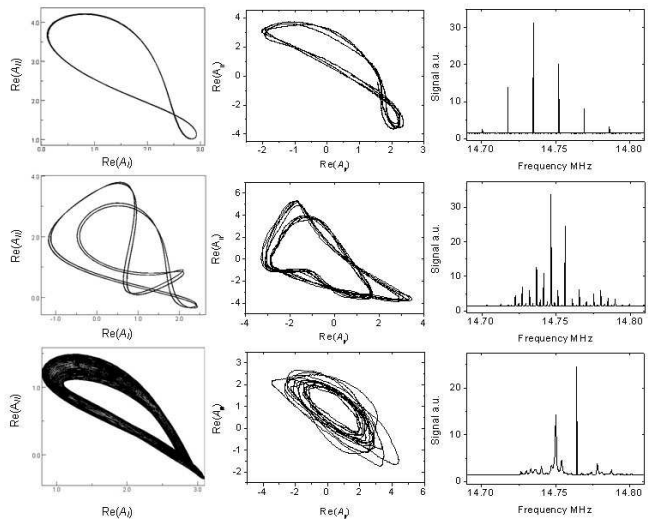}%
\caption{Complex dynamics of two strongly coupled nanomechanical resonators.
The three rows correspond to different input parameters (drive frequencies and
amplitudes). The first column shows the theoretical calculation of the phase
portrait, the second shows its experimental measurement, and the third column
shows the corresponding experimental power spectrum of the optical measurement
near one of the drive frequencies.}%
\label{Fig_4}%
\end{center}
\end{figure}
Examples of simulated and measured amplitude modulation dynamics for
three different sets of drive parameters are shown in Fig. 4 for a
second device with different parameters to the one used for
Figs.~1-3. The first column
shows examples of numerically calculated phase portraits of $\operatorname{Re}%
A_{II}$ versus $\operatorname{Re}A_{I}$ obtained by solving the time dependent
equations (\ref{Eq_Amplitudes}). Phase portraits measured experimentally are
shown in the second column \footnote{The full phase space is four dimensional,
$\operatorname{Re}A_{I},\operatorname{Re}A_{II},\operatorname{Im}%
A_{I},\operatorname{Im}A_{II}$; we plot a two dimensional projection.}. These
are obtained using homodyne down-conversion of the transduced mechanical
signals from both modes,
which are then read by independent oscilloscope channels.
We also perform wider frequency band spectrum analyzer
measurements shown in the third column. Since the nature of the dynamics is
sensitive to the precise values of the system parameters, the drive strength
and frequencies in the experiment are slightly adjusted from the values
corresponding to the theoretical plots to produce comparable phase portraits.

The top row of Fig.~4 shows a relatively simple example where the motion is
periodic but the $A_{I}-A_{II}$ trajectory forms a small figure-of-eight loop.
The dynamics can be roughly understood in terms of transitions between states
with the first mode at large amplitude and the second mode at small amplitude
and \textit{vice versa}.
The spectrum of the measured mechanical signal shows satellite peaks
corresponding to the anharmonic amplitude modulation.

As the parameters of the system are changed, we have observed period doubling
or quadrupling in the amplitude modulation, where the phase trajectory takes
two or four revolutions in order to complete the cycle \cite{Strogatz}. An
example of period quadrupling is shown in the middle row of Fig.~4. The
complicated loop structures are visible in both theoretical and experimental
trajectories, and the spectrum reveals amplitude modulation peaks at
frequencies that correspond to double and quadruple periods. Period doubling
transitions are often associated with chaotic dynamics \cite{Strogatz}, and
indeed for other parameter values we observe chaos in our coupled
nanoelectromechanical system and in the theoretical model, as shown in the
bottom row of Fig.~4. The evidence for the chaotic dynamics in the experiment
is the broad band component to the spectrum (evident in the shoulders to the
amplitude modulation peaks), and a phase portrait trajectory that does not
form a closed loop \footnote{Previous experiments that suggested chaos in a
nanomechanical system \cite{Blick} demonstrated a complex behavior of the response
at one of the drive frequencies as that
parameter was swept, rather than complex dynamics for fixed system parameters
as shown in our work.}. The theoretical model shows a similar phase portrait.

Our detailed study of two elastically coupled, independently driven,
nanomechanical beam resonators reveals complex nonlinear dynamics with a
number of potential applications. For example, driving one of the modes can be
used to tune the effective nonlinearity of the other mode. This can be used to
significantly increase the dynamic range of the resonator by quenching the
effective nonlinearity. In a first approximation, the motion of one mode
couples quadratically to the resonance frequency of the other mode, a
phenomenon that has been proposed for quantum nondemolition measurements in
nanomechanical systems. For larger vibration amplitudes spontaneous
oscillations of amplitude modulation develop, at a frequency determined by the
resonator ring-down time. These oscillations show period doubling and chaos
characteristic of strongly nonlinear systems. The full range of complex
dynamics investigated is quantitatively reproduced by theory.
Our success at predicting and subsequently observing quite delicate features
of the nonlinear dynamics is strong evidence that the nonlinearity and
coupling in arrays of nanomechanical devices can be quantitatively understood
and controlled.

\begin{acknowledgments}
Many of the ideas leading to this work were developed in
collaboration with Ron Lifshitz supported by the  U.S.-Israel
Binational Science Foundation (BSF) through Grant No. 2004339. We
thank Matt Matheny for many useful discussions.
\end{acknowledgments}


\begin{thebibliography}{99}                                                                                               %


\bibitem {PhWorld}M.\ L.\ Roukes, Phys.\ World \textbf{14}, 25 (2001).

\bibitem {LCreview}For a review of nonlinear dynamics in nanomechanical systems see R.\ Lifshitz and M.\ C.\ Cross, Reviews of Nonlinear
Dynamics and Complexity, \textbf{1}, 52 (2008)

\bibitem {Phil}Y.\ T.\ Yang, C.\ Calegari, X.\ L.\ Feng, K.\ L.\ Ekinci, and M.\ L.\ Roukes, Nano Lett.\ \textbf{6}, 583 (2006).

\bibitem {RugarSpin}D.\ Rugar, R.\ Budakian, H.\ J.\ Mamin, and B.\ W.\ Chui,
Nature \textbf{430}, 329 (2004).

\bibitem {ClelandCov}A.\ N.\ Cleland and M.\ L.\ Roukes, Nature
\textbf{392}, 160 (1998).

\bibitem {Matt}M.\ D.\ Lahaye, O.\ Buu, B.\ Camarota, and K.\ C.\ Schwab,
Science \textbf{304}, 74 (2004).

\bibitem {DynamicRange}H.\ W.\ C.\ Postma, I.\ Kozinsky, A.\ Husain, and M.\ L.\ Roukes,
Appl.\ Phys.\ Lett.\ \textbf{86}, 223105 (2005).

\bibitem{BuksRoukes}E.\ Buks and M.\ L.\ Roukes, Europhys.\ Lett.\ \textbf{54}, 220 (2001).

\bibitem {Lifshitz}R.\ Lifshitz, M.\ C.\ Cross, Phys.\ Rev.\ B \textbf{67}, 134302 (2003).

\bibitem {Parametric}Y.\ Bromberg, M.\ C.\ Cross and R.\ Lifshitz, Phys.\ Rev.\ \textbf{E73}, 016214 (2006).

\bibitem{Eyal}E.\ Kenig, R. Lifshitz, and M.\ C.\ Cross, preprint, 2008, arXiv:0808.3589v1 [nlin.PS]

\bibitem {D-NEMS}S.\ C.\ Masmanidis, R.\ B.\ Karabalin, I.\ Vlaminck, G.\ Borghs, M.\ R.\ Freeman, and M.\ L.\ Roukes,
Science \textbf{317}, 780 (2007).

\bibitem {Carr}D.\ W.\ Carr and H.\ G.\ Craighead, J.\ Vac.\ Sci.\ Technol.\ \textbf{15}, 2760-2763 (1997).

\bibitem {Goldstein}H.\ Goldstein, C.\ Poole, and J.\ Safko, \emph{Classical Mechanics},
3rd edition, (Addison-Wesley, new York, 2001)

\bibitem {RassulThesis}R.\ B.\ Karabalin, PhD dissertation, Caltech (2008).

\bibitem {Nayfeh}A.\ H.\ Nayfeh, D.\ T.\ Mook, \emph{Nonlinear Oscillations},
(Wiley, New York, 1979).

\bibitem {Braginsky}V.\ B.\ Braginsky, F.\ Y.\ Khalili, \textit{Quantum
Measurement}, (Cambridge University Press, Cambridge 1995).

\bibitem {PhysTodayQM}K.\ C.\ Schwab, M.\ L.\ Roukes, Physics Today, July
(2006), p36.

\bibitem {Strogatz}For a general discussion of period doubling and chaos see
S.\ H.\ Strogatz, \textit{Nonlinear Dynamics and Chaos}, (Addison-Wesley, New
York 1994).

\bibitem {Blick}D.\ V.\ Scheible, A.\ Erbe, R.\ H.\ Blick, and G.\ Corso, Appl.\ Phys.\ Lett.\ \textbf{81}, 1884 (2002).
\end{thebibliography}
\end{document}